\documentclass[a4paper,prl,final,superscriptaddress,twocolumn]{revtex4-1}
\usepackage[utf8]{inputenc}
\usepackage[english]{babel}
\usepackage[draft]{graphicx}
\usepackage{amsmath}
\usepackage{amssymb}
\usepackage{mathrsfs}
\usepackage{textcomp}
\usepackage{tabularx}
\usepackage{color}
\usepackage{array}
\usepackage{float}

\definecolor{darkblue}{rgb}{0.1,0.2,0.6}
\definecolor{darkred}{rgb}{0.8,0.1,0.2}
\usepackage[colorlinks,citecolor=darkblue,linkcolor=darkred,urlcolor=darkblue]{hyperref}

\begin{document}

\title{Signatures of van Hove singularities probed by the supercurrent in a graphene - hBN superlattice}
\author{D.I. Indolese}
\email{david.indolese@unibas.ch}
\affiliation{Department of Physics, University of Basel, Klingelbergstrasse 82, CH-4056 Basel, Switzerland}
\author{R. Delagrange}
\email{raphaelle.delagrange@unibas.ch}
\affiliation{Department of Physics, University of Basel, Klingelbergstrasse 82, CH-4056 Basel, Switzerland}
\author{P. Makk}
\affiliation{Department of Physics, University of Basel, Klingelbergstrasse 82, CH-4056 Basel, Switzerland}
\affiliation{Department of Physics, Budapest University of Technology and Economics and Nanoelectronics Momentum Research Group of the Hungarian Academy of Sciences, Budafoki ut 8, 1111 Budapest, Hungary}
\author{J.R. Wallbank}
\affiliation{National Graphene Institute, University of Manchester, Manchester, M13 9PL, UK}
\author{K. Wanatabe}
\affiliation{National Institute for Material Science, 1-1 Namiki, Tsukuba 305-0044, Japan}
\author{T. Taniguchi}
\affiliation{National Institute for Material Science, 1-1 Namiki, Tsukuba 305-0044, Japan}
\author{C. Sch\"{o}nenberger \footnote{author to whom correspondence should be addressed}}
\affiliation{Department of Physics, University of Basel, Klingelbergstrasse 82, CH-4056 Basel, Switzerland}

\begin{abstract}

The moir\'e superlattice induced in graphene by the hexagonal boron nitride substrate modifies strongly the bandstructure of graphene, which manifests itself by the appearance of new Dirac points, accompanied by van Hove singularities. In this work, we present supercurrent measurements in a Josephson junction made from such a graphene superlattice in the long and diffusive regime, where that the supercurrent depends on the Thouless energy. We can then estimate the specific density of states of the graphene superlattice from the combined measurement of the critical current and the normal state resistance. The result matches with theoretical predictions and highlights the strong increase of the density of states at the van Hove singularities. By measuring the magnetic field dependence of the supercurrent, we find the presence of edge currents  at these singularities. We explain it by the reduction of the Fermi velocity associated with the flat band at the van Hove singularity, which suppresses the supercurrent in the bulk while the electrons at the edge remain less localized, resulting in an edge supercurrent. We attribute this different behavior of the edges to defects or chemical doping.

\end{abstract}

\maketitle
The combination of graphene with different other 2D materials is a powerful means to engineer its electronic properties \cite{Ponomarenko2013,Woods2014}, for instance by inducing spin-orbit coupling \cite{Wang2015a, ZWang2016,Ren2016,Volkl2017,Zihlmann2018,Wakamura2018} or exchange interactions \cite{Wang2015,Leutenantsmeyer2017}. In particular, if graphene is placed on top of a hexagonal Boron Nitride (hBN) substrate, by aligning their crystallographic axes, a moir\'e superlattice is formed, which induces a periodic potential on a scale one hundred times larger than the interatomic distance in graphene leading to the modification of the bandstructure of graphene \cite{Yankowitz2012}. The wavelength $\lambda$ of the periodic potential defines new Brillouin zone boundaries, where satellite Dirac points (sDPs) may appear \cite{Park2008a, Yu2014}. In addition, van Hove singularities (vHSs) emerge in the density of states (DOS) at saddle points in the bandstructure due to the flattening of the arised minibands. These vHSs are encountered at much lower energy than in standard graphene, where they are only reachable by chemical doping \cite{McChesney2010}. Because the DOS diverges and charge carriers of different sign coexist, a rich physics is expected at the vHS, such as the formation of charge/spin-density waves \cite{Kiesel2012,Li2010} or unconventional superconducting pairing mediated by electron-electron interaction \cite{McChesney2010}. Moreover, the Chern number is predicted to change from subband to subband \cite{Brown2018}, leading to valley Hall effect and topological edge current when the DOS is gapped at the main Dirac point (mDP) \cite{Gorbachev2014,Zhu2016}. Very recently an intrinsic superconducting and a Mott insulating phase have been found in twisted bilayer graphene superlattices \cite{Cao2018a,Cao2018}.

Graphene-hBN superlattices \cite{Handschin2017, Ponomarenko2013, Woods2014} and the induced vHSs \cite{Li2010, Brihuega2012, Kim2016} have been widely studied with normal metal leads, but only few experiments have focused on the consequences of this rich physics for the Josephson effect. The investigation of the non-dissipative current induced in a non-superconducting system using a Josephson junction (JJ) geometry is  a powerful tool to investigate its physical properties, since the supercurrent is sensitive to the transport regime (ballistic/diffusive) \cite{Li2016,Ke2016,Borzenets2016,Murani2017,Nanda2017}, interactions \cite{vanDam2006,DeFranceschi2010} and to the current distribution within the sample. For example, Josephson interferometry has been used recently to detect the presence of edge current in quantum spin Hall systems \cite{Hart2013,Pribiag2015} and in graphene where edge current was observed close to the Dirac point due to guided wave states \cite{Allen2015} or, in bilayer graphene, due to the opening of a gap using an electric field \cite{Zhu2016}. In this last article, edge current in a graphene/hBN superlattice at the mDP is reported, where it is claimed that a gap opens due to sublattice symmetry breaking \cite{Hunt2013,Woods2014}. In contrast to these previous works, we investigate the supercurrent over the full range of energy, in order to probe the superlattice bandstructure.

We investigate the superconducting transport in long, diffusive JJs made from graphene/hBN superlattice and show that the supercurrent carries in this transport regime the signature of its very specific bandstructure, in particular of the vHSs. First, by measuring both the normal state resistance and the supercurrent and taking advantage of the diffusive regime, we estimate the DOS of the JJs, which is then compared to theoretical calculations for a moir\'e superlattice. Further, we extract the current distribution in the sample as a function of the charge carrier density from the magnetic field dependence of the supercurrent  and show that edge currents appear at the vHSs, where the DOS diverges. We show that this edge current corresponds to a suppression of the supercurrent in the bulk, associated with the reduction of the Fermi velocity at the singularity that globally localizes the electrons. This suppression is not observed in the edges, probably because some edge defects or doping reduce the influence of the superlattice.

\begin{figure}[htb]
	\centering
	\includegraphics[width=\columnwidth]{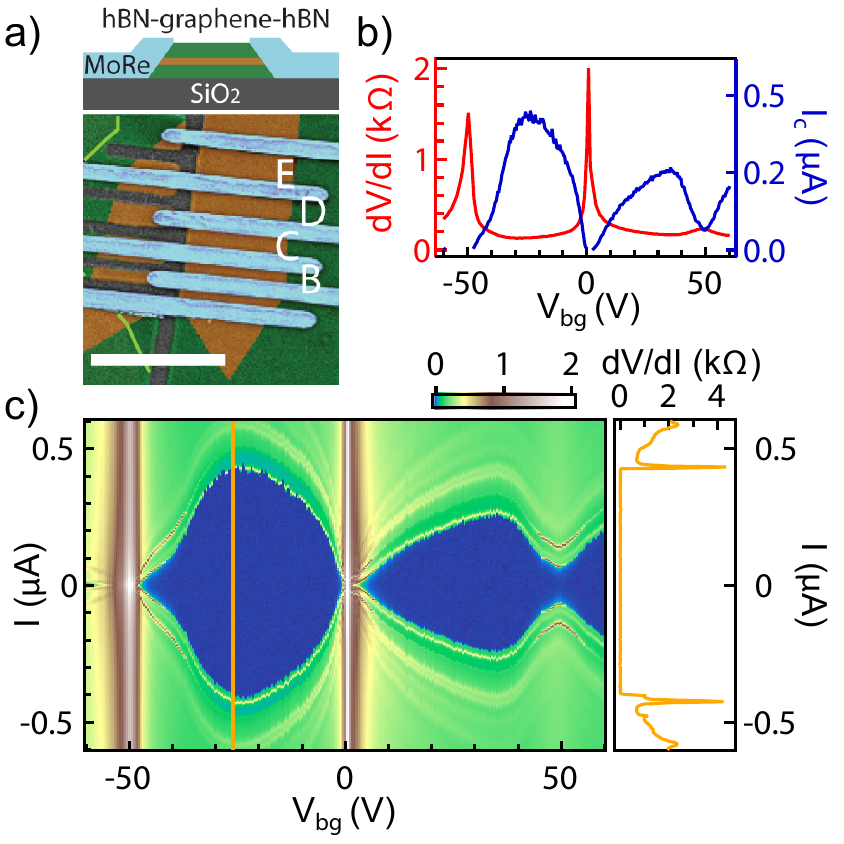}
	\caption{a) Top: Schematic side view of the stack. Bottom: False colored SEM image of the device. The graphene (brown) is encapsulated in between hBN (green) and contacted with MoRe (blue). The white scale bar corresponds to $5\mathrm{~\mu m}$. b) Normal state resistance $R_N$ (red) and critical current $I_c$ (blue) as a function of gate voltage $V_{g}$  for junction D.  c) Differential resistance as a function of $V_{g}$ and DC current bias $I$ for junction D. 
		 Right: line cut at the indicated gate voltage ($V_g=-26\,V$).}
	\label{fig:Fig_1}
\end{figure}

The measured sample is a hBN-graphene-hBN stack, where one of the hBN is aligned with the graphene. The heterostructure is contacted with one dimensional superconducting edge-contacts \cite{Wang2013}. We fabricated the electrodes by co-sputtering of MoRe (1:1) chosen for its large critical magnetic field (8T) as well as its high critical temperature (7K) \cite{Calado2015, Amet2015}. Several JJs are realized in the same stack with different lengths $L$ from $0.45$ to $1\mathrm{~\mu m}$ and a width of $W=3\mathrm{~\mu m}$ (Fig.\ref{fig:Fig_1}a). All electronic transport measurements are performed in a dilution refrigerator at a base temperature of 70\,mK. 

Since the critical field of MoRe is too large to suppress the superconductivity by applying a magnetic field, we estimated the junction resistance $R_N$ from the quasiparticle current measured when the JJ is voltage biased with $|eV|>2\Delta_{MoRe}$, with $\Delta_{MoRe}=1.3$\,meV the superconducting gap of MoRe, estimated from multiple Andreev reflections (see Supplemental Material \cite{SuppI}). The measurement is performed in a two terminal configuration, such that $R_N$ contains the resistance of the graphene channel $R_G$ together with the contact resistance $2R_c$ ($R_N=R_G+2R_c$).

In the four junctions investigated, we observe an enhancement of resistance around the mDP ($V_g=0V$) and in addition around $V_g=\pm50V$, corresponding to a charge carrier density $n_0=\pm 3.3\times10^{12}\mathrm{~cm^{-2}}$ (see Fig.\ref{fig:Fig_1}b). These additional resistance maxima are attributed to sDPs in the bandstructure and are clear evidence of a superlattice \cite{Ponomarenko2013,Hunt2013}. From the value of $n_0$, we estimate the misalignment angle between the graphene and the hBN lattice to be around 0.7$^\circ$. Note as well that no gap opening is observed at the mDP (Supplemental Material \cite{SuppI}). The analysis of the gate dependent resistivity shows that all junctions are in the diffusive regime, where the mean free path is smaller than the junction length $L$.

 We first measure the critical current $I_c$, defined as the maximal current that can be passed through the junction. To do so, we current bias the sample and measure the differential resistance as a function of bias current $I$ and gate voltage as shown in Fig.\ref{fig:Fig_1}c for sample D (see Supplemental Material \cite{SuppI} for samples B, C, E). The switching from the zero resistance state to the normal resistance state is detected as a sharp transition at $I=I_c$, as presented in the right panel of Fig.\ref{fig:Fig_1}c and plotted as a function of $V_g$ on Fig.\ref{fig:Fig_1}b. No hysteresis was observed between the retrapping  and switching current, indicating that the JJ is in the overdamped regime. At the first order, $I_c$ is inversely proportional to $R_N$, and is thus strongly reduced at the Dirac points, beyond the resolution of the measurement. $I_c$ is globally smaller for electron doping ($V_g>0$) than for hole doping. This reduction of the supercurrent can be attributed to a p-doping of the graphene by the MoRe, leading to the formation of a p-n junction between the metal contacts and n-doped graphene.  Note that in previous works n-doping of the contacts was observed \cite{Calado2015, Borzenets2016}. This difference may be attributed to the work functions of graphene and MoRe which are almost the same \cite{Yu2009,Nasa1970}.

If the time $\tau$ spent by the electrons in the junction is short compared to $\hbar/\Delta$, in an ideal JJ the product of the normal state resistance with its critical current is expected to be proportional to the superconducting gap $\Delta$ \cite{Tinkham}. But if $\tau$ exceeds $\hbar/\Delta$, then the relevant energy scale becomes the Thouless energy such that $eR_NI_c=\alpha E_{th}$, with $\alpha$ a constant that depends on the transport regime (ballistic or diffusive) \cite{Dubos2001,Li2016,Ke2016,Borzenets2016}. The four junctions we investigated are in this regime, since the superconducting coherence length $\xi_S<200\,$nm$<L$ (Supplemental Material \cite{SuppI}). In agreement with Refs. \cite{Li2016,Ke2016}, we assume that the finite reflection probability at the contacts leads to an increase of $\tau$ such that it can be included as a reduction of $\alpha$.  Combining the expression of the Thouless energy $E_{th}=\frac{\hbar D}{L^2}$ with the Einstein relation $L/WR_G=De^2\times DOS$, we find that the DOS as a function of the charge carrier density $n$ can be determined from the measurement of both $R_N$ and $I_c$:

\begin{equation}
\label{equ:DOS}
DOS(n)=\alpha\frac{\hbar}{R_N(n) R_G(n) e^3 L W I_c(n)}.
\end{equation}

Note that this formula involves $R_G$, which is obtained by subtracting the contact resistance $R_c$ from the measured resistance $R_N$. 

The DOS expected in the graphene-hBN superlattice was calculated using the methods described in Ref. \cite{Wallbank2013}. The DOS on the hole side vHS is quite robust to small changes of the moir\'e parameters used in the theoretical model, while on the electron side it depends significantly on their choice. We chose here parameters similar to those extracted in Ref. \cite{Lee2016}, adapted to $\theta$ = 0.7$^\circ$, but slightly modified to produce a vHS on the electron side similar to previous measurements \cite{Yu2014}.

To compare our data with the theoretical calculated DOS, we have to make several assumptions: (i) the measurement of the critical current $I_c$ is not affected by the finite temperature, (ii) the coefficient $\alpha$ is constant over the investigated gate range and (iii) the contact resistance $R_c$ is constant respectively for electron and hole doping. For the electronic temperature $T=100\,$mK, we estimate that hypothesis (i) is correct for measured supercurrents higher than $30\mathrm{~nA}$ \cite{SuppI}, which excludes the gate regions around the mDP and the  sDP at the hole side from the analysis. Concerning (ii), Refs. \cite{Li2016} and \cite{Ke2016} have shown that $\alpha$ is indeed constant for a long diffusive graphene JJ, even if the measured value of 0.1-0.2 is substantially lower than the one expected for a SNS junction \cite{Dubos2001}. (iii) is the strongest hypothesis, since $R_c$ can actually depend on $V_g$ and can vary within a factor of two around the mDP \cite{Russo2010, Xia2011}, but we believe that even a gate dependent contact resistance would not change the qualitative picture outlined below.

\begin{figure}[h]
	\centering
	\includegraphics[width=\columnwidth]{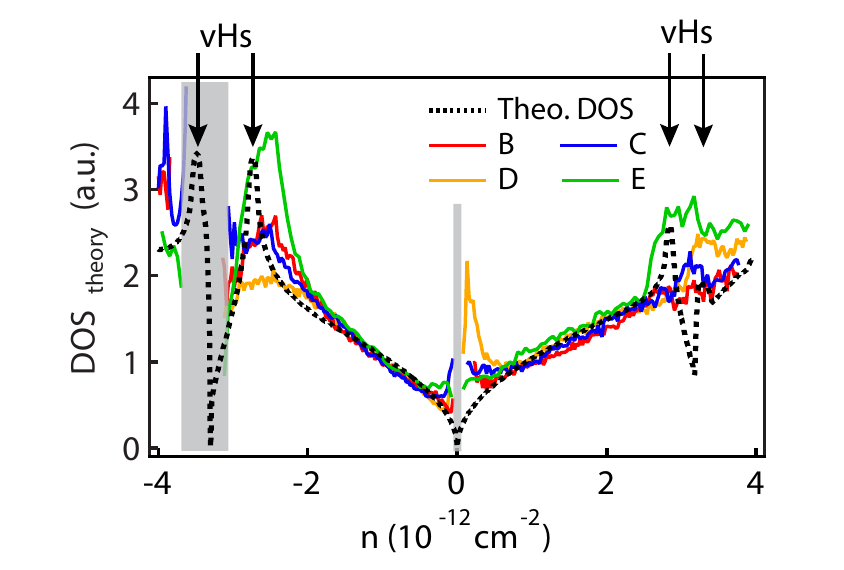}
	\caption{Density of state estimated from measured $R_N$ and $I_c$ (Eq.\ref{equ:DOS}) (in red, blue, yellow and green respectively for B, C, D and E) compared to a calculation for $\theta=0.7^\circ$ (black), as a function of the charge carrier density. The moiré superlattice parameters (defined in Ref \cite{Lee2016}) used to produce the theoretical DoS are $U^+_0$=8.5\,meV, $U^+_1$=-8.5\,meV, $U^+_3
$=-14.7\,meV.}
	\label{fig:Fig_2}
\end{figure}

Then, by taking $R_c$ and $\alpha$ as fitting parameters, we are able to reproduce the calculated DOS using Eq.\ref{equ:DOS} for $\alpha\in [0.3,0.8]$ and $R_c\approx40-160\mathrm{~\Omega}$ (see Supplemental Material \cite{SuppI}). The result is plotted in Fig.\ref{fig:Fig_2}. For the four junctions, the DOS matches the calculation and reproduces DOS over a large gate range. As  theoretically expected, the superlattice features are less pronounced on the electron side.  As a whole, despite some strong assumptions and some uncertainty in the precise value of the contact resistance, we show that  the combined measurement of $I_c$ and $R_N$ allows to estimate the DOS, providing information about the specific bandstructure of the superlattice. In particular, we see a clear signature of the vHSs, which was not explicitly present in either $R_N$ or $I_c$.\\

It can be noted that the vHS at negative $V_g$ is more pronounced for junction B, C and E than for junction D. In order to understand this discrepancy, we look now into the current distribution in junction D (see Supplemental Material \cite{SuppI} for junction C) using the so called Josephson interferometry, which consists in measuring the magnetic field dependence of the supercurrent \cite{Zhu2016, Allen2015}. The magnetic field induces a relative phase shift between the different supercurrent paths, which then interfere in the leads and generate an interference pattern related to the Fourier transform of the supercurrent distribution in the sample.

\begin{figure}[h]
	\centering
	\includegraphics[width=\columnwidth]{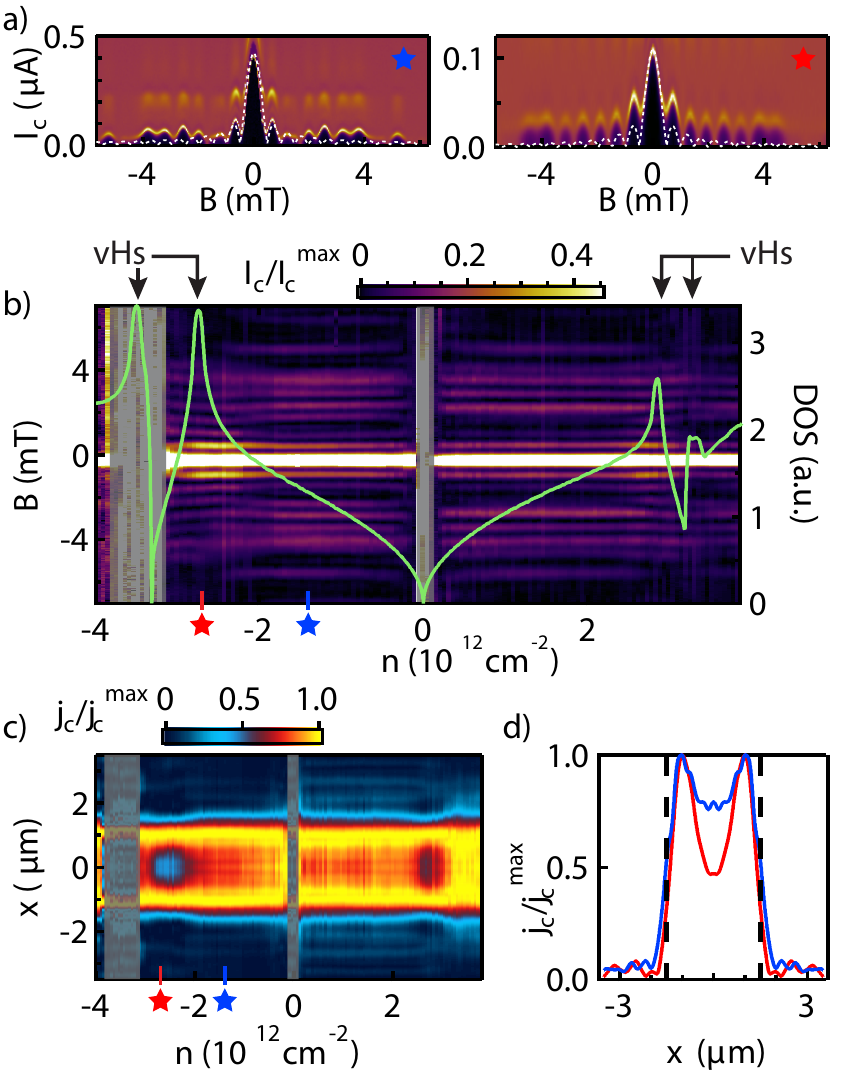}
	\caption{a) Differential resistance as a function of current bias and magnetic field at $n_1=-1.4\times10^{12}\,$cm$^{-2}$ (blue star) and $n_2=-2.7\times10^{12}\,$cm$^{-2}$ (red star). The white dashed line is the Fraunhofer pattern expected for a homogeneous current density. b) Normalized critical current as a function of magnetic field $B$ and carrier density $n$ measured in junction D, superimposed with the calculated DOS in green. The position in $n$ of the measurement shown is indicated by stars. c) Calculated current density as a function of carrier density $n$ and position along the contacts. d) Linecuts of panel c) at $n_1$ (blue) and $n_2$ (red). The black dashed lines indicate the sample edges at $\pm 1.5\,\mu$m.}
	\label{fig:Fig_3}
\end{figure}

Typical interference patterns are represented in Fig. \ref{fig:Fig_3}a and compared to the Fraunhofer interference pattern, expected for a homogeneous current distribution \cite{Cuevas2007} and a sinusoidal current phase relation as measured for graphene JJs \cite{Nanda2017,Bretheau2017a}. At $V_{g}=-20\,$V ($n_1=-1.4\times10^{12}\,$cm$^{-2}$), between the mDP and the vHS, the interference pattern matches a Fraunhofer pattern for the first few lobes, with a periodicity consistent with the sample dimensions taking the finite field penetration into the superconductor into account \cite{Allen2015}. At slightly higher fields ($B>\pm1.5\,$mT), one can see some missing lobes and a non-vanishing supercurrent, indicating that the current is not perfectly homogeneous. The pattern at $V_{g}=-40\,$V ($n_2=-2.7\times10^{12}\,$cm$^{-2}$), close to the vHS, is strikingly different, since the first lobes and the central peak are of comparable amplitude, which is an indication of SQUID-like current distribution \cite{Hart2013} and a enhanced edge current.

In order to understand the gate dependence, we measure the interference pattern over the gate range from minus to plus $60\,V$. For that, we bias the sample with a linearly increasing current, at a rate $0.17\,$A/s. The critical current is obtained from the time at which the junction turns normal, averaged 200 times. The interference pattern can then be plotted as a function of gate voltage (Fig. \ref{fig:Fig_3}b). In order to compare the shape of the interference patterns, for each $V_g$ the supercurrent is normalized by its maximum value, $I_c(B=0)$ for each gate voltage. Note that this kind of measurement cannot detect currents smaller than a few tens of nA, given by $V_t/R_N$ with $V_t$ the threshold voltage for the switching to normal conducting state. 

We can distinguish two different regimes for the interference pattern: far from the vHSs, the interference pattern is gate independent and similar to the one described in Fig.\ref{fig:Fig_3}a left. In contrast, around both vHSs, the pattern is similar to Fig.\ref{fig:Fig_3}b right, where the side lobes become more prominent. The effect is stronger for hole doping, where the vHS is more pronounced. 

To be more quantitative, we calculate the current distribution in the sample by the inverse Fourier transformation of the interference pattern for each $V_g$. The exact procedure is described in the Supplement Material \cite{SuppI} and follows the ansatz given in \cite{Hart2013,Allen2015}. The full map of the current density $j_c$ as a function of $V_g$ is shown Fig.\ref{fig:Fig_3}c, where $j_c$ was normalized by the maximal current density of each trace in $n$ similar to Fig.\ref{fig:Fig_3}b. Two representative distributions are plotted in Fig.\ref{fig:Fig_3}d for $n_1$ (blue) and $n_2$ (red), showing that in the whole sample a part of the current accumulates on the edges, and that the proportion of edge to bulk current is significantly larger at the vHS.

 \begin{figure}[h]
	\centering
	\includegraphics[width=0.8\columnwidth]{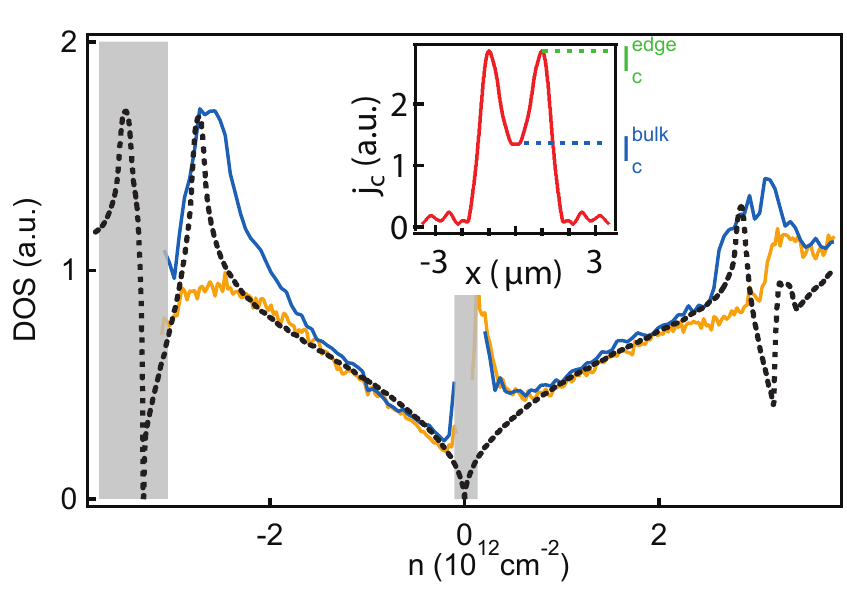}
	\caption{DOS of the bulk (blue) as a function of charge carrier density $n$ in junction D, estimated from the bulk  current (see inset) and the total current (yellow). Inset: current distribution at the vHS at negative charge carrier density.}
	\label{fig:Fig_4}
\end{figure}

From the non-renormalized map of the supercurrent distribution, we are able to extract separately the gate dependence of the supercurrent on the edges of the sample ($I_c^{edge}$) and in the bulk ($I_c^{bulk}$) defined as shown in Fig.\ref{fig:Fig_4}. In order to elucidate the nature of the edge current, we use the same procedure as for Fig.\ref{fig:Fig_2} to estimate the DOS of the bulk. For that, we use $I_c^{bulk}$ instead of $I_c$ and the same resistances $R_N$ and $R_c$ (assuming that the normal state resistance is dominated by the bulk). 
The result is shown in Fig.\ref{fig:Fig_4}. We find a very good agreement between the DOS extracted from $I_c^{bulk}$ (blue) with the theoretically determined DOS (dotted). In particular, the vHS is now better reproduced than using the total current $I_c$ (plotted in yellow for comparison), meaning that the edge current doesn't carry the signature of the vHS. On the other hand, due to the flat band at the vHS, the Fermi velocity is expected to be globally reduced in the superlattice. This tends to localize the electron by increasing the traversal time $\tau$ of the electron in the junction and leads therefore to a reduction of the supercurrent. This localization acts weaker on the electrons at the edges, which leads to an increased edge to bulk current ratio at the vHS.

 We performed the same measurement and data analysis for junction C (see Supplemental Material \cite{SuppI}). There, the presence of edge current is not observed as in sample D and the DOS extracted using Eq.\ref{equ:DOS} exhibits a clear pronounced increase at the vHS. These two facts suggest that, in sample C, the edges are more affected by the superlattice potential than in sample D, and show that both measurements of current distribution and DOS from $R_N$ and $I_c$ are consistent and complementary. 

It remains to understand why the edges are behaving differently from the bulk in sample D. One can rule out the hypothesis of topological edges states due to the valley Hall effect at a gap opening (as proposed in \cite{Brown2018} and measured in \cite{Gorbachev2014}), because the current at the edges appears far from any bandcrossing. It has been shown that edge current can be induced as well by guided-wave electronic states due to the band bending at the sample edges \cite{Allen2015}, but only close to the Dirac point, where the edge potential is unscreened. Our measurement would be more consistent with previous works reporting edge current induced by electrostatic or chemical doping of the edges \cite{Dou2018a, Woessner2016, Panchal2014, Zhu2016}. This may induce disorder that can affect the superlattice potential, such that the vHS may be smoothened \cite{Lherbier2008}. This alteration could originate from the exposure of the graphene edge to ambient condition during the fabrication or from the contamination during the reactive ion etching used to shape the sample. 

In conclusion, we demonstrate in this work that the supercurrent carries the signature of the modified graphene bandstructure by the moir\'e superlattice. First, from the combined measurement of the normal resistance and the critical current and taking advantage of the diffusive regime, we estimate the DOS in the sample and find a very good qualitative agreement with the DOS calculated theoretically. In addition, Josephson interferometry reveals the presence of a gate dependent edge current in junction D and its portion is strongly enhanced at the vHSs. By estimating the DOS for the bulk, we show that the edges are less affected by the superlattice potential, probably due to edge disorder or chemical doping. We then attribute the edge current to the lowering of the Fermi velocity in the bulk associated with the flat band at the vHs. 

Beyond the specificities of our sample, this work demonstrates the possibility of observing edge supercurrent in graphene far from the DPs, shedding a new light on Josephson interferometry experiments. 

\section*{Acknowledgement}

DI and RD contributed equally to this work.
The authors thank C. Handschin and S. Zihlmann for their help in the lab and helpful discussions. We thank as well S. Goswami, Chuan Li, H. Bouchiat, S. Guéron and V. Fal'ko for insightful discussions.

This work has received funding from ERC project TopSupra (787414), the European Union Horizon 2020 research and innovation programme under grant agreement No 696656 (Graphene Flagship), the Swiss National Science Foundation, the Swiss Nanoscience Institute, the Swiss NCCR QSIT, Topograph, ISpinText FlagERA network and from the OTKA PD-121052 and OTKA FK-123894 grants. P.M. acknowledges support from the Bolyai Fellowship. Growth of hexagonal boron nitride crystals was supported by the Elemental Strategy Initiative conducted by the MEXT, Japan and the CREST (JPMJCR15F3), JST.

\end{document}


\maketitle

\section{Additional information about the sample} 

\subsection{Fabrication}

A dry pick-up technique developed by Zomer et al. \cite{Zomer2014} was applied to encapsulate the graphene in between two hBN flakes. Graphene and hexagonal Boron Nitride (hBN) flakes were exfoliated (Nitto tape) onto p-doped silicon wafer with a thermally grown 300$\,nm$ thick SiO$_2$ layer. The used hBN crystals were grown by K. Watanabe and T. Taniguchi and the graphite source was HOPG from "HQ-graphene".  The stack was annealed in a H$_2$/N$_2$  atmosphere for 50\,min at 300\,$^\circ$C. The contacts were defined by e-beam lithography using 300\,nm thick PMMA. After cold development (at 0$^\circ$C) in IPA/DI-water (7:3), the graphene was contacted using reactive ion etching process with a gas mixture of SF$_6$, O$_2$ and Ar to open an access to the encapsulated graphene, which then was contacted in a quasi four terminal configuration with sputtered MoRe (1:1) leads (thickness 50\,nm). The liftoff was performed in warm (50\,$^\circ$C) acetone. In an additional lithography step the MoRe was contacted by Ti/Au (5\,nm/50\,nm) and bond pads were defined. In the end the width of the transport channel was defined using the same reactive ion etching process as mentioned before.

\subsection{Determination of the superlattice wavelength}

\begin{figure}[h]
	\begin{center}
		\includegraphics[height=5cm]{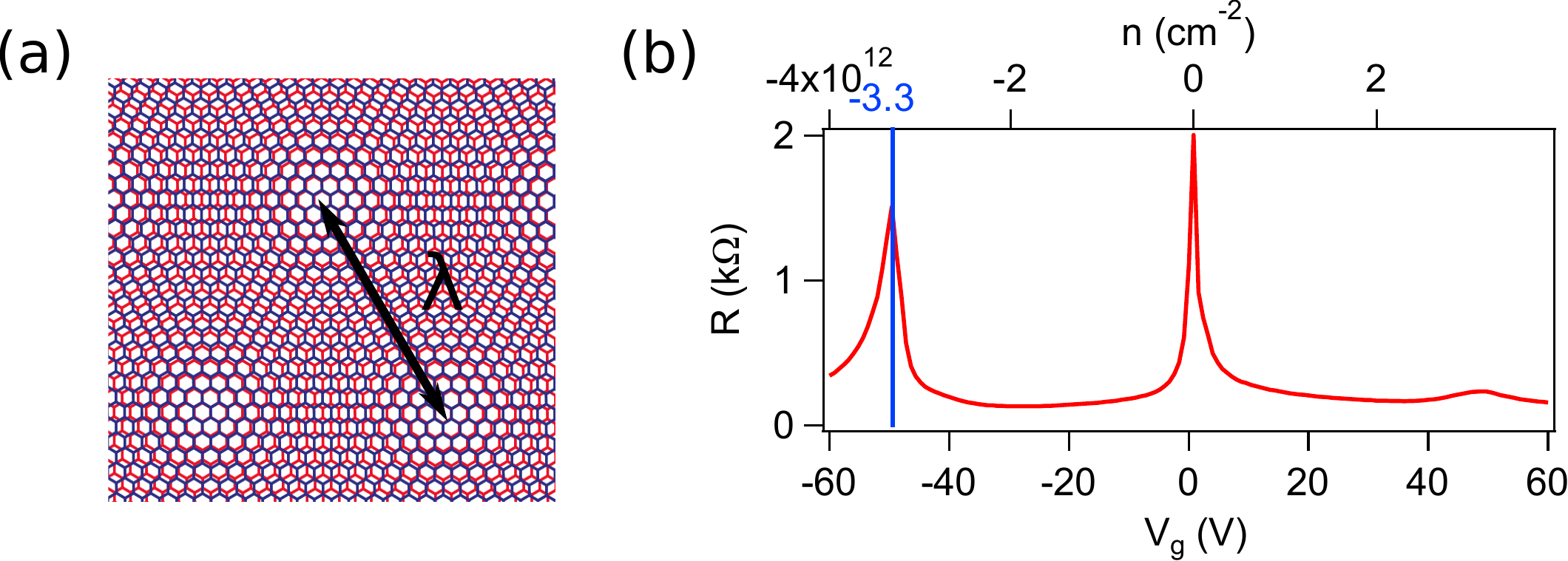}
	\end{center}
	\caption{(a) Schematics of a moiré superlattice formed by alignment of a graphene layer on h-BN. The wavelength of the new hexagonal superlattice formed is called $\lambda$. (b) Normal state resistance measured in sample D (see main text) as a function of gate voltage, converted to charge carrier density $n$ with a capacitance model.  }
	
	\label{DP_position}
\end{figure}

When the crystallographic axis of a graphene layer is aligned (or almost) with the one of a hexagonal Boron Nitride (h-BN) layer, because the two lattices have almost the same lattice constant (with a mismatch of $\delta = 1.8\%$ \cite{Yankowitz2012}), a moir\'e superlattice formes (see fig. \ref{DP_position} (a)). This hexagonal superlattice is characterised by its wavelength $\lambda$ and the corresponding wavevector $|G|=\frac{4\pi}{\sqrt{3}\lambda}$ \cite{Ponomarenko2013}.

Secondary Dirac points are expected to arise at the new Brillouin zone boundaries, at $\vec{k}$ vectors such that $k=|\vec{k}|=|G|/2$. In a 2D material, k can be expressed as a function of the charge carrier density by $k=\sqrt{\pi n}$, such that the wavelength can be obtained from the position of the second Dirac peak $n_{SDP}$: $\lambda=\sqrt{\frac{4\pi}{3n_{SDP}}}$. In our sample, $n_{SDP}=-3.3.10^{12}\mathrm{~cm^{-2}}$ is extracted from the normal state resistance measurement of fig. \ref{DP_position}, giving $11.3\pm0.1\mathrm{~nm}$.

 The maximum value of $\lambda$ is of the order of $14\mathrm{~nm}$, corresponding to a perfect alignment of the lattices, and decreases with the misalignement angle. This misalignement angle $\theta$, expressed in radian, can be estimated from the following formula \cite{Wallbank2013}:
 \begin{equation}
G=\frac{4\pi}{\sqrt{3}a}\sqrt{\delta^2+\theta^2}
 \end{equation}
 with the lattice constant $a=2.46\mathring{A}$. In our sample, we get $\theta\approx 0.7^o $.

\subsection{Temperature dependence of the normal state resistance}

\begin{figure}[h]
	\begin{center}
		\includegraphics[height=4.5cm]{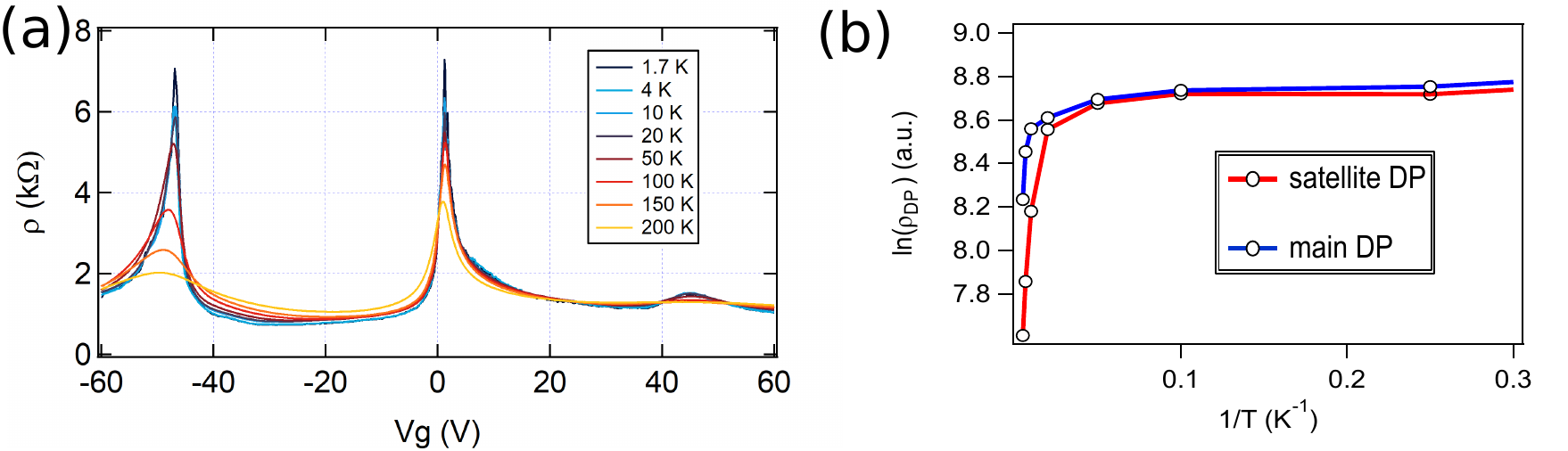}
	\end{center}
	\caption{(a) Temperature dependence of the resistivity as a function of the gate voltage, measured in sample C. (b) Arrhenius plot: resistivity at the left secondary Dirac point (red) an at the main Dirac point (blue), as a function of 1/T and in log scale. If there would be a thermally activated gap, it should be a line.}
	\label{T-dep}
\end{figure}

In some monolayer graphene h-BN superlattice, a gap opening has been predicted and observed \cite{Wallbank2013,Woods2014,Dou2017}, leading presumably to edge currents \cite{Zhu2017}. In order to determine if there is a gap opening in the density of states of our sample, we measure the resistivity as a function of temperature, represented in fig. \ref{T-dep}. One can see that, below $100\mathrm{~K}$, the resistivity only slightly varies. Note as well that the value of the resistivity is not as high as one would expect if there would be a gap \cite{Woods2014}.

If there is a gap, it should be equal to twice the thermal activation energy $E_A$ extracted from an Arrhenius plot. Namely, by plotting the logarithm of the resistivity as a function of $1/T$, where $T$ is the temperature, one expects a linear behavior with a slope of $E_A/k_B$. From fig. \ref{T-dep} (b), this linear behavior is not observed in our sample, except at large temperature ($T>100 K$) corresponding to an activation energy of 110 K, and thus a gap of 220 K. The extracted energy has to be compared with the disorder energy at the DP. This disorder manifests itself in charge puddles, which appear in our sample at charge carrier densities smaller then $3.10^{10}\mathrm{~cm^{-2}}$, corresponding to a energy of $20\mathrm{~meV}$. It is thus possible that there exists a gap in the system, maybe induced by the superlattice, which would be smeared out by disorder at low temperature. If so, it anyway doesn't affect the measurements of the main text, which are only performed below 100\,mK.

\subsection{Quantum Hall effect}

\begin{figure}[h]
	\begin{center}
		\includegraphics[height=5.4cm]{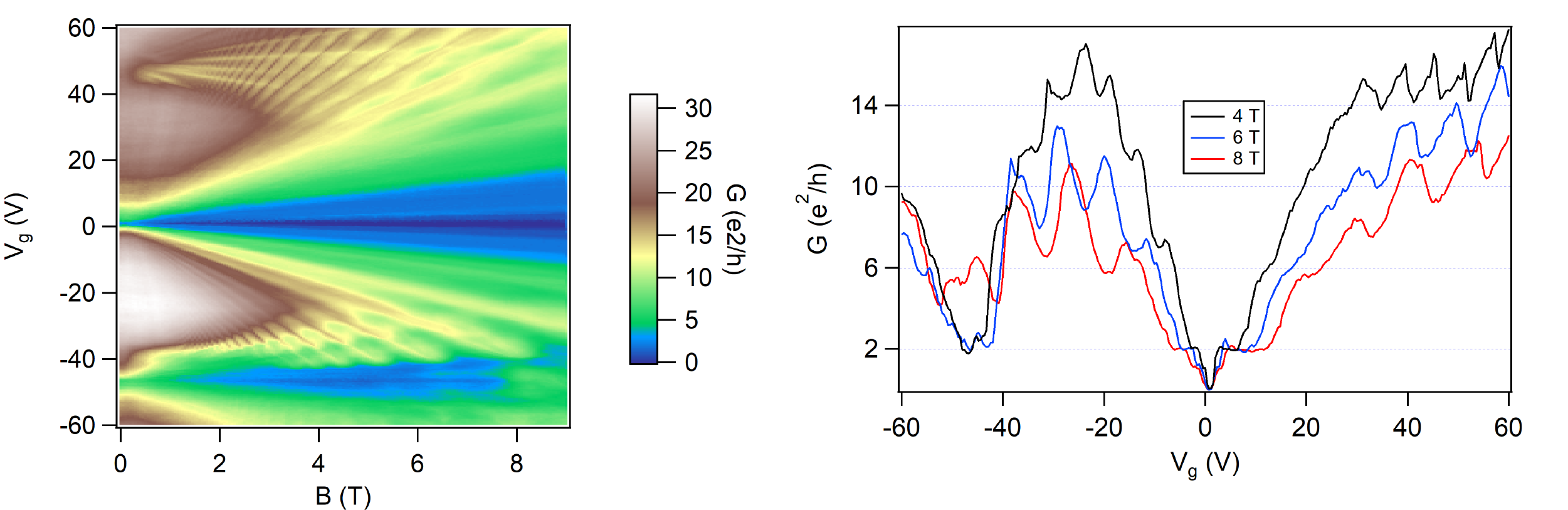}
	\end{center}
	\caption{Left: differential conductance G as a function of the magnetic field for various gate voltage. It is measured in sample E after reshaping to an aspect ratio close to 1. }
	\label{QHE}
\end{figure}

After the measurements described in the main text, the sample has been reshaped using a SF6 etching recipe. The new width, $1.2\mu m$ gives for sample E an aspect ratio close to 1, which is more appropriate for observing quantum Hall plateaus in a two terminal geometry \cite{Abanin2008}. Fig. \ref{QHE} shows the differential conductance $G=dI/dV$ of sample E as a function of the gate voltage and the magnetic field at a temperature of $1.8 K$. This measurement, showing a clear plateau at $2e^2/h$, confirms that the measured sample is a monolayer graphene. 

One can also see a well defined plateau at $e^2/h$ and faintly defined ones between 2 and $6e^2/h$ at 8 T. This shows that some degeneracies are lifted, especially for hole doping (where the sample is cleaner)\cite{Young2012}. This breaking of degeneracy at rather low field in a not-so-clean sample could be attributed to the superlattice \cite{Hunt2013,Yang2016}. The spacing between the split Landau levels is then four time smaller, which may be the reason why the plateaus are blurred.

 The features observed at larger gate voltage are strongly affected by the superlattice band structure of the sample, their detailed investigation would require a Hall bar measurement and is far beyond the scope of this work.

\section{Electronic transport}

\subsection{Transport regime}

\subsubsection{Contact resistance and mean free path}

We estimated the mobility $\mu$ of the graphene device and the contact resistance $R_c$ at the interface with MoRe by fitting the conductivity $\sigma$ as a function of carrier density $n$ using 

\begin{equation}
\label{eq:u_fit}
\sigma^{-1}=(ne \mu +\sigma_0)^{-1}+\rho_c,
\end{equation} 

where $e$ is the electron charge,  $\sigma_0$ the residual conductivity at the charge neutrality point and $\rho_c$ corresponds to a carrier density independent contribution to the resistivity from contact resistance and short range scattering. Note that this procedure is valid only in the vicinity of the charge neutrality point and becomes irrelevant close to the satellite Dirac point (ie for $|V_g|>20 V$), which limits the range of the fit and thus the reliability of the results. Still, in order to give an order of magnitude of the quality of the sample, $\mu$ and $R_c$ are listed for the four samples in Tab.\ref{table}. 

To decide if our samples are diffusive or not, we have to estimate their mean free path. For that, we relate the diffusion coefficient $D$ to the density of states $DOS$ through the Einstein relation $\sigma=D e^2 DOS(E_F)$. Then, assuming that the density of state is the one of a graphene monolayer (which is approximately true for a superlattice around the main DP), we get (independently of the value of the Fermi velocity):
\begin{equation}
l_{mfp}\equiv \frac{2D}{v_F}=\frac{L\sqrt{\pi}\hbar}{R_N W e^2 \sqrt{n}}
\end{equation} 

The values of the mean free paths for each junction is estimated in table \ref{table} for a charge carrier density of $n=10^{16}m^{-2}$ (when two values are given, they correspond to hole and electron doping respectively).

\begin{table}[h]
	\begin{center}
		\begin{tabular}{c||c|c|c|c|c|}
			Sample & L ($nm$) & $l_{mfp}(nm)$ & $\xi_s (\mu m)$ & $2R_c(\Omega)$ &  $\mu (cm^2/Vs)$  \\
			\hline
			B & 450 &  60/35 & 0.1 &315/350 &  7'500/4'000 \\
			C & 640 & 115/85 & 0.15 &137/193 &  14'000/12'000 \\
			D & 820 & 190/120 & 0.2 &80/92 &  30'000/11'000 \\
			E & 1000 &  230/150 & 0.2 &102/137 & 37'000/22'000 
		\end{tabular}
		\caption{Summary of the transport properties of the four junctions measured (mean free path and superconducting coherence length) together with $\mu$ and the contact resistance $R_c$ , the fitting parameters obtained using Eq.\ref{eq:u_fit}.}
	\end{center}
	\label{table}
\end{table}

From this table, the four samples are diffusive ($L>l_{mfp}$). The $l_{mfp}$ is proportional to $\sqrt{n}$ and therefore it depends only weakly on $n$ for high doping. Moreover, in the vicinity of the satellite Dirac points, the mean free path may decrease again, such that the sample stays in the diffusive regime all over the investigated range of gate voltage. The mobility is unusually low for a hBN stack, and the smaller is the junction, the more diffusive it is. This may be due to the close  proximity of the MoRe contacts, which deteriorates the quality of the sample.

\subsubsection{Long or short junction?}

It is important to know if our junctions are in a short or a long junction regime regarding the superconducting proximity effect. In a short junction, one can neglect the phase acquired by the Andreev pair during the propagation in the junction compared with the phase acquired in the Andreev reflection. This short junction regime corresponds to
\begin{equation}
L\ll\xi_S=\frac{\hbar v_F}{\pi \Delta}\approx 200\mathrm{~nm}
\end{equation} 
with $\Delta=1.3\mathrm{~meV}$ (value estimated from multiple Andreev reflections, see below) and $v_F=1\times10^6\mathrm{~m/s}$ (in a superlattice, it may be smaller, so this value is an upper bound). In our samples, $L>450nm$, so we are in the \textbf{long junction regime}.

In the diffusive limit, the coherence length is actually $\xi_S=\sqrt{\hbar D/\Delta}$, which gives similar order of magnitude (see table \ref{table}, where it has been estimated as well with $v_F=1\times10^6\mathrm{~m/s}$).

\subsection{Supercurrent measurement}

\subsubsection{Additional information about the estimation of the DOS from $R_NI_c$ product}

\paragraph{$R_N$ and $I_c$ for the four junctions investigated}

\begin{figure}[htbp]
	\centering
	\includegraphics[height=7cm]{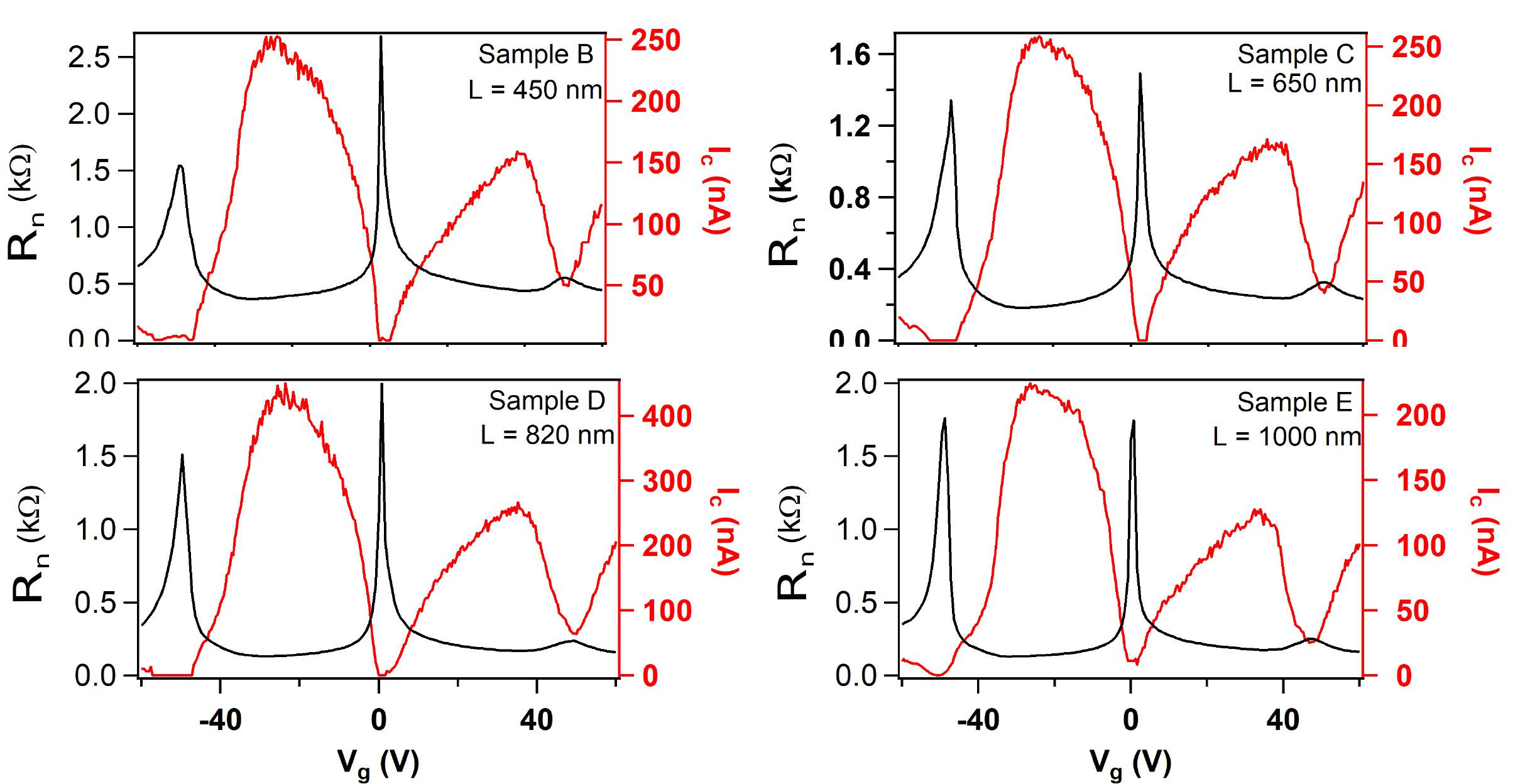}	
	\caption{Normal state resistance $R_N$ and critical current $I_c$ as a function of the gate voltage $V_g$ for the four junctions investigated.}
	\label{fig:RnIc}
\end{figure}

We show on fig. \ref{fig:RnIc} the normal state resistance and the critical current as a function of gate voltage for the four junctions measured. The resistance plotted is the measured one, including the contact resistance.

\begin{figure}[h]
	\centering
	\includegraphics[height=5.5cm]{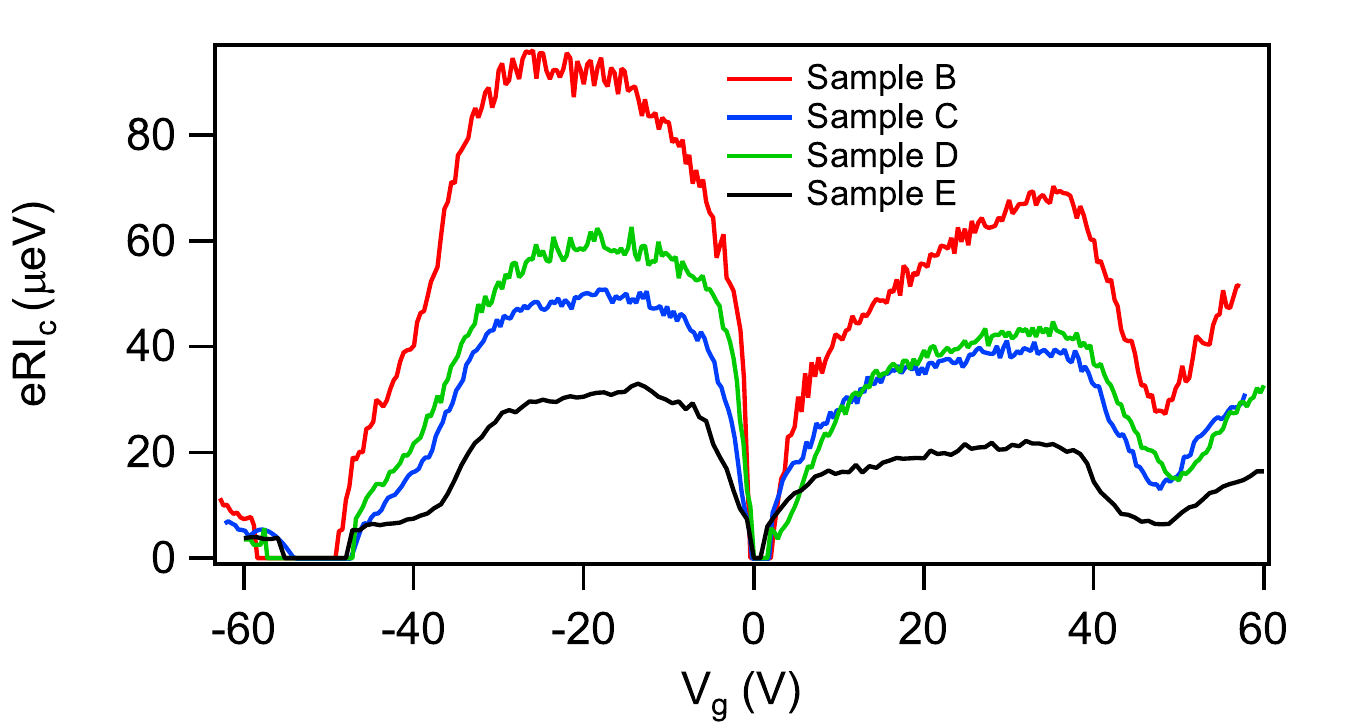}	
	\caption{Product of the normal state resistance $R_N$ with the critical current $I_c$ as a function of the gate voltage $V_g$ for the four junctions investigated.}
	\label{fig:RnIc2}
\end{figure}

Fig. \ref{fig:RnIc2} shows the product of the normal state resistance $R_n$ with the critical current $I_c$.

\paragraph{Determination of the DOS}

As explained in the main text, in a long and diffusive Josephson junction, the density of state can be expressed as a function of the critical current and of the normal state resistance through:
\begin{equation}
\label{equ:DOS}
DOS=\alpha\frac{\hbar}{(R_N-2R_c) R_N e^3 L W I_c}.
\end{equation}

$\alpha$ is defined such that $eR_nI_c=\alpha E_{th}$ and is not known a priori in graphene \cite{Li2016,Ke2016}. The contact resistance $R_c$ is not known precisely as well. These two parameters are thus taken as gate-independent fitting parameters that we adjust manually to obtain a good agreement with the calculated DOS. The obtained values are given in table \ref{table2}.

\begin{table}[h]
	\begin{center}
		\begin{tabular}{c||c|c|c|}
			Sample & L ($\mu m$)  & $2R_c(\Omega)$ &  $\alpha$  \\
			\hline
			B & 0.45 &  275-320 &  0.75-0.7 \\
			C & 0.64 & 115-150 &  0.4 \\
			D & 0.82 & 80-100 &  0.45 \\
			E & 1 & 75-110 & 0.3-0.27 
		\end{tabular}
		\caption{Values of $\alpha$ and of the contact resistance $R_c$, the fitting parameters used to adjust our data to the calculated DOS (see main text). These values are known with an uncertainty of the order of 10\%.}
	\end{center}
	\label{table2}
\end{table}

Note that this calculated DOS is obtained by setting some parameters that we cannot determine experimentally and may be altered by the disorder in the sample. For these reasons, added to the fact that our measurement of supercurrent is not reliable close to the Dirac point, the agreement is more qualitative than quantitative. Consequently, we estimate the uncertainty on the given values of $\alpha$ and $R_c$ to be of the order of 10\%.

The values of $R_c$ are consistent with the contact resistance estimated from the gate dependence of the resistivity (see above). The values of $\alpha$ are consistent as well with the ones that can be found in the literature \cite{Li2016,Ke2016}. \\

\subsubsection{Switching/critical current}

In a finite temperature measurement, one doesn't measure exactly the critical current of the Josephson junction but rather what is called a switching current, resulting from the thermal activation of the switching to the non-superconducting state. This point is important for the estimation of the density of states as described in the main text, where we assumed that the measurement represented the real critical current.\\

To quantify this thermal effect, we use the resistively and capacitively shunted junction (RCSJ) model. It consists in modelling the Josephson junction as a perfect Josephson element with a sinusoidal current-phase relation $I=I_c\sin(\varphi)$ (for graphene, this is a reasonable assumption), in parallel with a resistance and a capacitor. The whole system is biased by a current $I$ and $V$ is the voltage across it.	

\begin{figure}[h]
	\begin{center}
		\includegraphics[height=4cm]{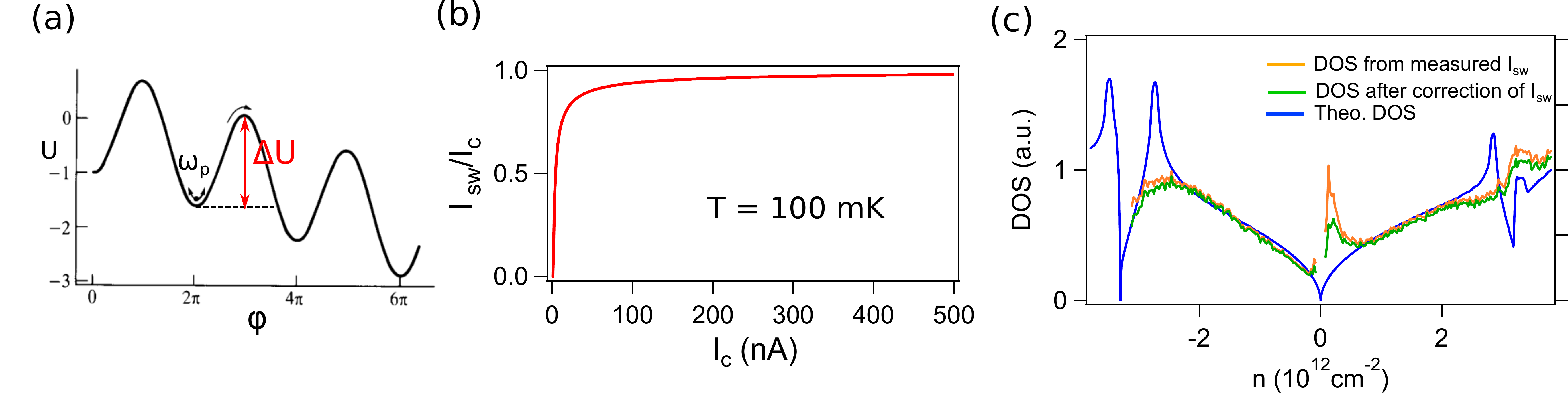}
	\end{center}
	\caption{(a) The phase dynamics is equivalent to the movement of a particle in the effective potential U, and submitted to a frictional force (see text). This potential is a cosine, tilted proportionally to the current I. (b) Ratio between the estimated switching current and the critical current as a function of the critical current. This ratio corresponds to the current at which the switching probability reaches 0.5. (c) DOS extracted from the measurement of $R_N$ and $I_c$ for sample D (see main text). In orange, the value of $I_c$ used is the one directly measured, the switching current $I_{sw}$. In green, the switching current has been corrected using the ratio plotted on fig. (b) to obtain the real $I_c$. In blue is recalled the DOS theoretically expected.   }
	\label{RCJS}
\end{figure}

Using the Josephson relation $\frac{d\varphi}{dt}=\frac{2eV}{\hbar}$ and Kirchhoff's current law, we write \cite{Tinkham}:
\begin{equation}
\frac{d^2\varphi}{dt^2}=-\omega_p^2\sin(\varphi)+\omega_p^2\frac{I}{I_c}-\frac{\omega_p}{Q}\frac{d\varphi}{dt}
\end{equation}
with $\omega_p=\sqrt{2eI_c/\hbar C}$ and $Q=\omega_p RC$.
This is formally  the equation of motion of a particle, whose position is given by $\varphi$, in an effective potential (fig. \ref{RCJS}):
\begin{equation}
\label{potential_RCJS}
U(\varphi)=-E_J\cos(\varphi)-\frac{\hbar I}{2e}\varphi
\end{equation}
and subjected to a frictional force $\frac{\hbar}{2e}^2\frac{1}{R}\frac{d\varphi}{dt}$, with $E_J=-\frac{\hbar}{2e}I_c$. 

At zero temperature $T=0$, while $I<I_c$, the phase is trapped in a local minimum of potential, where it oscillates at frequency $\omega_p$. When the current is increased to $I=I_c$, there is no barrier preventing the phase from increasing leading to a rapid onset of voltage $V=\frac{\hbar}{2e}\frac{d\varphi}{dt}$ across the junction. The current becomes dissipative. 

At finite temperature, thermal activation allows the fictitious particle to leave its local potential minimum for $I<I_c$. This current at which the particle tunnels across the barrier while increasing I is called the switching current $I_{sw}$, and is always lower than the critical current. To estimate the ratio between the measured switching current and the "real" critical current, we calculate the barrier \cite{Fulton1976}:
\begin{equation}
\Delta U=\frac{\hbar I_c}{2e}\left(\frac{I}{I_c}(2\sin^{-1}\frac{I}{I_c}-\pi)+2\cos(\sin^{-1}\frac{I}{I_c})\right)
\end{equation}
According to Arrhenius law, the probability of switching (i.e. the probability of overcoming the barrier) is then $P_{sw}=e^{-\Delta U/k_BT}$ \cite{Clarke1988}. For each value of $I_c$, the value of the current needed to reach a switching probability of 0.5 has been extracted, and is plotted on fig. \ref{RCJS} (b) for an estimated electronic temperature of $100\mathrm{~mK}$ corresponding to our measurements. We see qualitatively that, if the critical current is higher than $30\mathrm{~nA}$, this ratio is roughly constant.

To figure out to what extent the temperature affects our conclusions, we plot on fig. \ref{RCJS} (c) a figure similar to fig. 2 of the main text, where we compare the DOS extracted from the measurement with and without taking into account the finite temperature. Far from the Dirac points, the difference between the two quantities is very tiny, below the uncertainty of the measurement. This is why, in the main text, we used the measured current without applying any correction and called it $I_c$. At the Dirac point, where the supercurrent cancels, temperature effect may explain the unexpected increase of the DOS, even though our simple model is not able to explain it fully.\\

Note that, in this qualitative estimation of the effect of finite temperature, we neglected the influence of quantum macroscopic tunnelling \cite{Devoret1985} and assumed that the sweep velocity of the bias current was infinitely slow.

\subsection{Determination of the superconducting gap of MoRe}

\begin{figure}[htbp]
	\centering
	\includegraphics[]{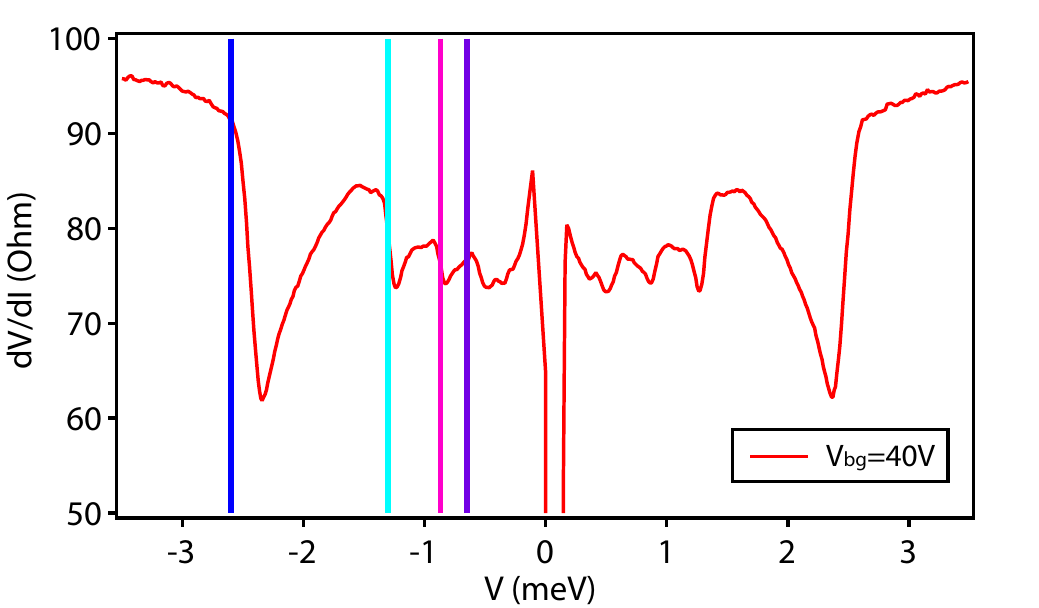}	
	\caption{Differential resistance plotted as a function of V$_{bias}$ at a V$_{bg}$=40\,V. The vertical blue line indicates the position of 2$\Delta$, while the other colored lines (light blue, pink and violet) are showing the position of the MAR at V$_{bias}=\Delta,~2\Delta/3\mathrm{~and~}\Delta/2$.}
	\label{fig:Vbias}
\end{figure}

The superconducting gap of MoRe was measured in a different graphene stack, which doesn't show any transport signatures of a superlattice. To do so, the differential resistance was measured with a standard lock-in technique in a quasi four terminals configuration as a function of voltage bias V$_{bias}$ and an ac-voltage amplitude of 20\,$\mu$V at 377\,Hz. A clear decrease of the differential resistance is observed at V$_{bias}$=2.6\,meV, that we attribute to twice the superconducting gap of MoRe $\Delta_{MoRe}$ and indicated by the dark blue line in Fig.\ref{fig:Vbias}. This value ($\Delta_{MoRe}=1.3\mathrm{\,meV}$) is of the same order as measured by Borzenets et al. \cite{Borzenets2016} ($\Delta$=1.2\,meV). Additional peaks of the differential resistance inside the gap arise due to multiple Andreev reflection MAR at V$_{bias}$ equal to $\Delta$ (light blue), 2$\Delta$/3 (pink) and $\Delta$/2 (violet). In the region around V$_{bias}$=0, the dV/dI drops to 0 because of a supercurrent flowing through the JJ as soon as V$_bias$/R$_L$<I$_c$m, where R$_L$ is the line resistance, since the finite value of R$_L$ (R$_L$=137\,$\Omega$) causes an effective current bias of the sample.

\section{Analysis of interference patterns: extraction of the current distribution}

In this section, we describe the procedure used to extract the current distribution $j_s(x)$ from the interference pattern of the critical current $I_c$ as a function of magnetic field $B$, following \cite{Dynes1971,Allen2015,Hart2014}. We choose the graphene sheet to lie in the x-y plane with contacts along x from -W/2 to W/2, a length of L and an out of plane magnetic field in z-direction.

\begin{figure}[htbp]
	\centering
	\includegraphics[height=4cm]{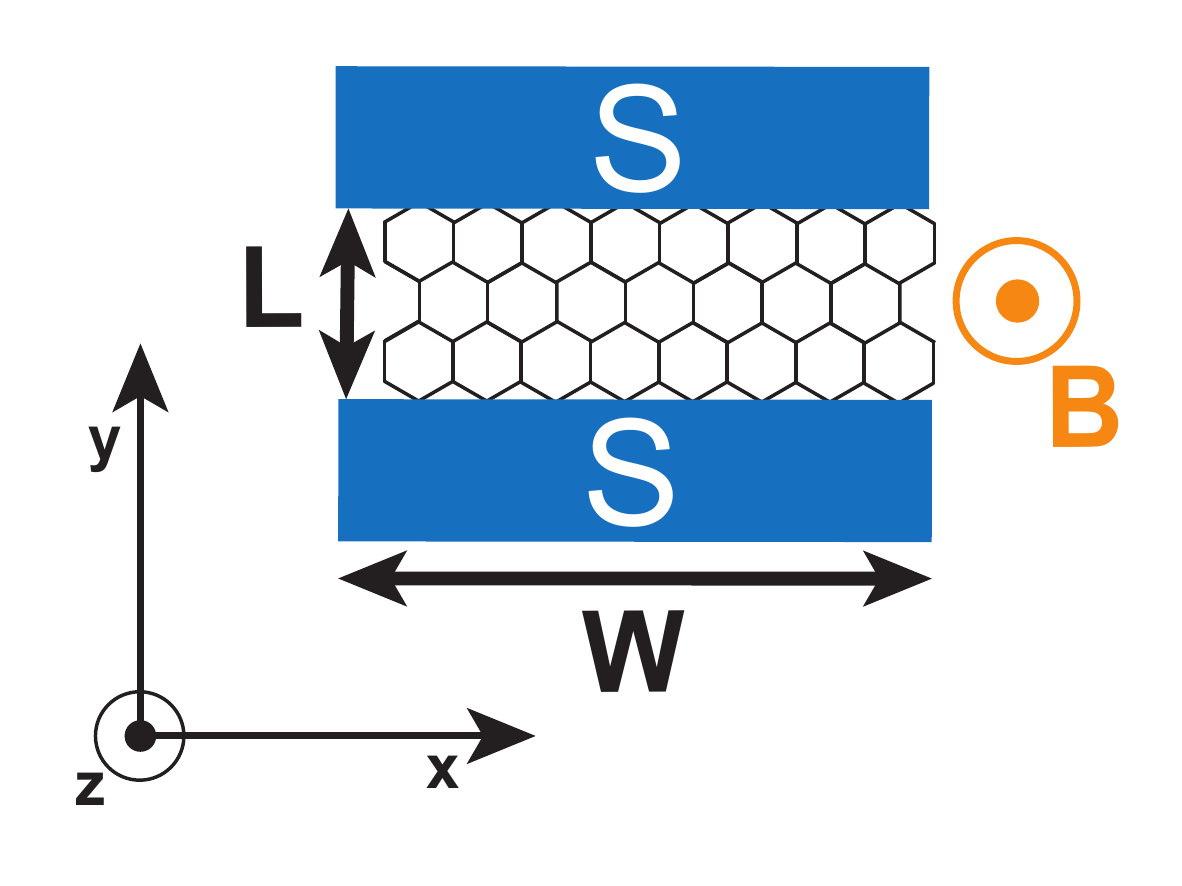}	
	\caption{Model of a graphene based Josephson junction with contacts in x direction, a supercurrent flowing in y. To measure the interference pattern, a perpendicular magnetic field in z direction is applied.}
	\label{fig:SampleModel}
\end{figure}

\subsection{Extracting $j$ from interference pattern}

\label{interf_patt_even}
We assume that the supercurrent in graphene can be described approximately with a sinusoidal current-phase relation, i.e.

\begin{equation}
\label{equ:cur-pha}
j_s(x)=j(x)\sin(\varphi(x)),
\end{equation}

where $j$ is the maximum supercurrent density and $\varphi$ the superconducting phase difference between the two superconductors. This is a very reasonable assumption if most of the transmission channels are not perfectly transmitted (which is more likely the case in our diffusive sample with non-perfect contact resistance) and is consistent with existing measurements in graphene \cite{Nanda2017,Bretheau2017a}.

If a perpendicular magnetic field is applied, two supercurrent paths flowing through the junction pick up an additional relative phase with respect to each other, which depends on the enclosed area by the two paths. Therefore, the superconducting phase at position x is given by the integration of the perpendicular magnetic field $B$ over the penetrated area $S$ added to a reference phase $\varphi_0 = \varphi(x=0)$ and is expressed as

\begin{equation}
\label{equ:phadiff}
\varphi(x)=\varphi_0+\frac{2\pi}{\phi_0}\int_{S}Bds=\varphi_0+\frac{2\pi\Phi(x)}{\phi_0},
\end{equation}

where $\phi_0=h/2e$ is the flux quantum and $\Phi=B*S(x)$ the magnetic flux. The area is given by $S(x)=(L+2\lambda)*x$, where the London penetration depth $\lambda$ is added twice to the junction length to take a finite penetration of the magnetic field in the contacts in to account. 

Therefore the total mediated supercurrent $J_s$ can be calculated by the integration along the sample width W.

\begin{equation}
\label{equ:Scur}
J_s(\beta,\varphi_0)=\int_{-W/2}^{W/2}dx j(x)\sin(\varphi_0+\beta x)=sin(\varphi_0)\int_{-W/2}^{W/2}dx j_e(x) e^{i\beta x}-i cos(\varphi_0)\int_{-W/2}^{W/2}dx j_o(x) e^{i\beta x},
\end{equation}

with $\beta = 2\pi B(L+2\lambda)/\phi_0$, $j_e(x)$ the even and $j_o(x)$ the odd part of $j(x)$. For a fixed magnetic field B, one can vary $\varphi_0$ to obtain the maximum current $J_c$ known as critical current of the JJ. We assume now an even current distribution, such that the problem simplifies and the measurable quantity $I_c$ can be express as

\begin{equation}
\label{equ:FT}
I_c(\beta)=\Bigl| J_c(\beta)\Bigl|= \Bigl| \int_{-W/2}^{W/2}dx j_e(x) e^{i\beta x} \Bigl|.
\end{equation}

\begin{figure}[h]
	\centering
	\includegraphics[width=\textwidth]{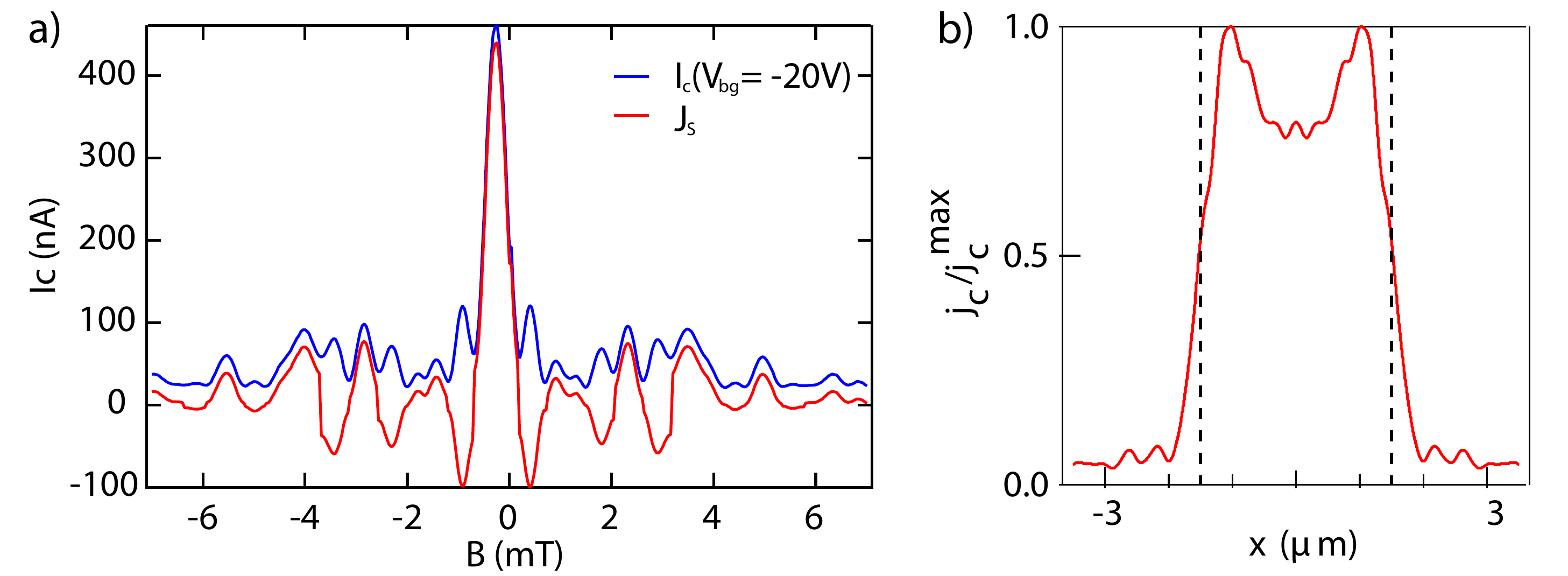}	
	\caption{a) Measurement of critical current  (blue) and reconstructed supercurrent (red) as a function of magnetic field. b) Calculated and normalized current density as a function of x. The black dashed lines are indicating the position of the junction edges along the contacts.}
	\label{fig:FFT_prep_D}
\end{figure}

Further, only the even part of the integral will be non zero leaving us with $J_c=\int_{-W/2}^{W/2}dx j_e(x)cos(\beta x)$, which can take negative or positive values (with a sign change at each zero crossing in the case of homogeneous current distribution ). The measurement, represented in blue in Fig.\ref{fig:FFT_prep_D}a, is actually $|J_c|$, such that we have to reconstruct the sign of $J_c$. This is done by inverting the sign of every second lobe of $I_c$ (see Fig.\ref{fig:FFT_prep_D}a). To prevent discontinuities, we subtracted a constant background from $I_c$ to shift its value to 0 for magnetic fields of 6\,mT. This background arises partly due to the used measurement method described in the main article. 

By calculating the inverse Fourier transform of $J_c$ the current density in real space can be determined as

\begin{equation}
\label{equ:invFT}
j(x)= \int_{\beta_{min}}^{\beta_{max}}d\beta J_c(\beta) e^{i\beta x}.
\end{equation}

The result of the described procedure is shown in Fig.\ref{fig:FFT_prep_D}b.

\subsection{Adding an odd component to the current distribution}

The non vanishing critical current at the minima of the interference pattern is an indication for a contribution of an odd part in the current distribution $j_o(x)$. Taking $j_o$ into account a new expression for $J_c$ is given by

\begin{equation}
\label{equ:odd}
J_c=\int_{-W/2}^{W/2}dx j_e(x)cos(\beta x)+i\int_{-W/2}^{W/2}dx j_o(x)sin(\beta x).
\end{equation}

\begin{figure}[h]
	\centering
	\includegraphics[width=0.6\textwidth]{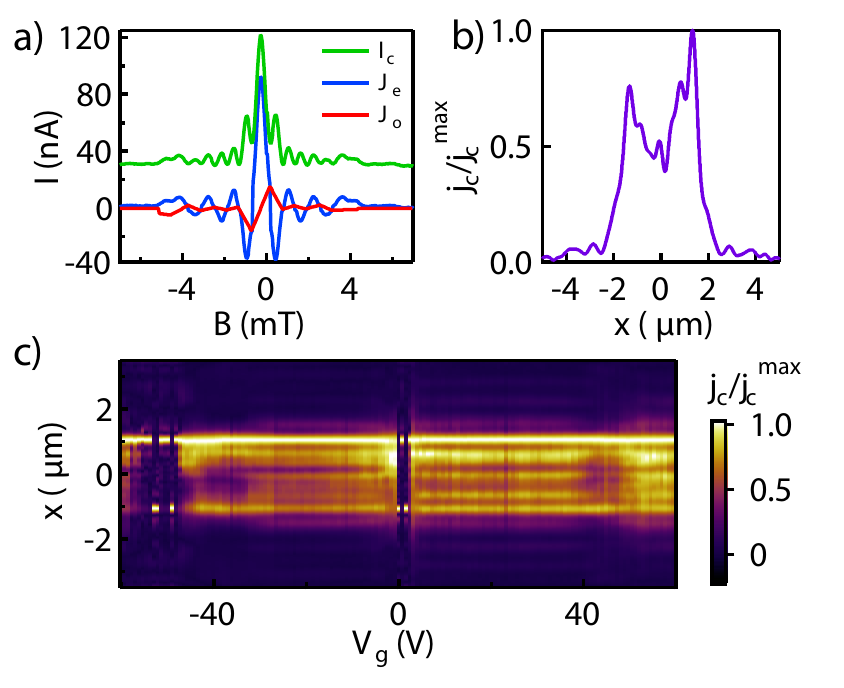}	
	\caption{a) Measured critical current (green) as a function of magnetic field at a gate voltage of -40\,V. In blue (red) the even (odd) part of the reconstructed supercurrent is shown. b) Current density at -40\,V as a function of x. c) Calculated current density including the asymmetric part as a function of back gate and position. An enhancement of the edge to bulk current density is observed around the van Hove singularities at $\approx\pm$40\,V.}
	\label{fig:FFT_asym}
\end{figure}

The critical current can now be written as $I_c=\lvert J_c\rvert=\sqrt{J_e^2+J_o^2}$, where $J_e$ and $J_o$ are the even and odd part of $J_c$. From the measured interference pattern we see the even part dominates most of the time. But, from Eq.\ref{equ:odd}, it follows that the odd part is dominating where $J_e$ vanishes, i.e. for small $I_c$. To reconstruct the odd contribution we followed the Ansatz in Ref.\cite{Hart2014} interpolating between the minima of $I_c(B)$ and flipping sign between each lobe (see Fig.\ref{fig:FFT_asym}a). The result of $j(x)$  is shown in Fig.\ref{fig:FFT_asym}b). We observe that it seems that one edge is contributing more to the supercurrent transport than the other. As described in the main article we are able to extract the density of states (DOS) of the bulk by extracting the dependence of the critical current with carrier density from the calculated current density as a function of gate shown in Fig.\ref{fig:FFT_asym}c. No qualitative difference in the resulting DOS was observed between the calculation with and without the odd component $j_o$. 

This analysis containing even more assumption, we rather limit ourselves to the even component, as describe in section \ref{interf_patt_even}.

\subsection{Interference pattern for sample C}

\begin{figure}[h]
	\centering
	\includegraphics[width=0.9\textwidth]{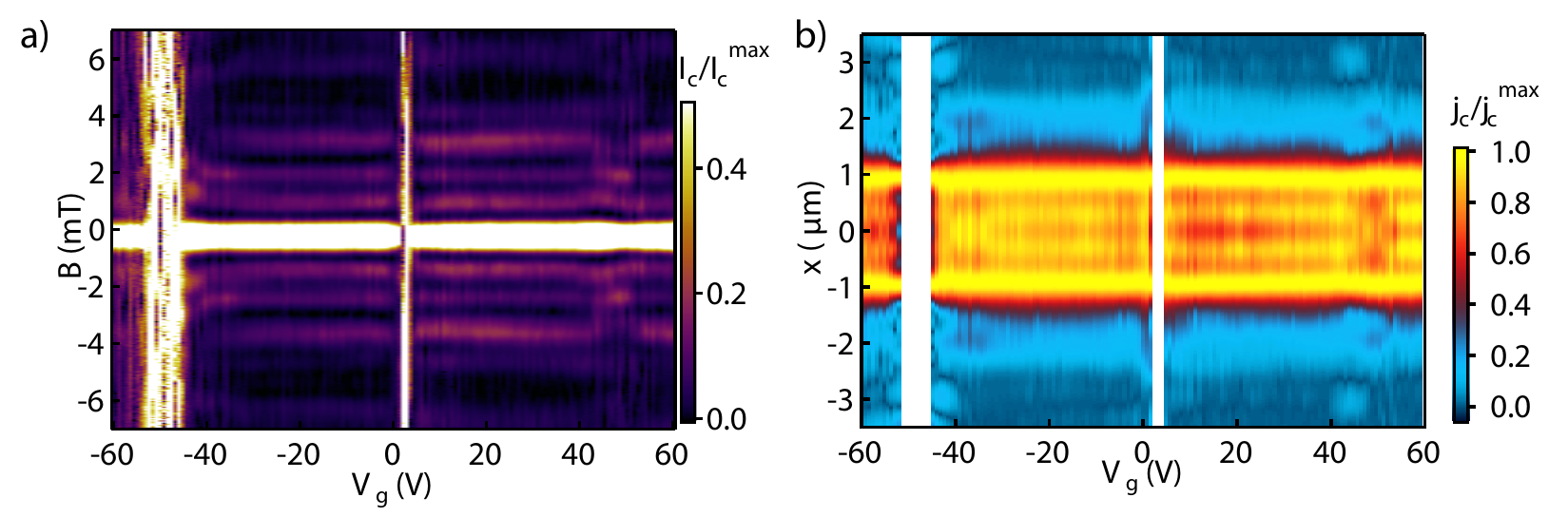}	
	\caption{a) Interference pattern of critical current of sample C as a function of gate voltage $V_g$ and magnetic field $B$. b) Calculated current density as a function of $V_g$ and position x for sample C. The white areas correspond to gates voltages, where the critical current was to small to be measured.}
	\label{fig:FFT_c}
\end{figure}

In addition to the current distribution of sample D (L=0.82\,$\mu$m), which is discussed in the main article, we also studied the one in sample C (L=0.64\,$\mu$m). The interference pattern as a function of gate voltage and magnetic field was measured and normalized as described for sample D (see Fig.\ref{fig:FFT_c}a). It shows a monotonous behaviour over the entire gate range. Small changes appear at the satellite Dirac points at positive gate voltage, which are probably due to the strongly reduced amplitude of the critical current, such that small features can not be resolved anymore due to the limited measurement resolution. The calculation of the current density does not carry any indication of a increased edge to bulk current ratio around the van Hove singularities. This behaviour is consistent with the observation that the estimated DOS from $R_N I_c$ of sample C probes the van Hove singularities without additional data processing like for sample D, where a finite current contribution by the edges had to be subtracted.
